# Dichotomy for Voting Systems

Edith Hemaspaandra*
Department of Computer Science
Rochester Institute of Technology
Rochester, NY 14623

Lane A. Hemaspaandra†
Department of Computer Science
University of Rochester
Rochester, NY 14627

April 15, 2005

**Abstract**

Scoring protocols are a broad class of voting systems. Each is defined by a vector $(\alpha_1, \alpha_2, \ldots, \alpha_m)$, $\alpha_1 \geq \alpha_2 \geq \cdots \geq \alpha_m$, of integers such that each voter contributes $\alpha_1$ points to his/her first choice, $\alpha_2$ points to his/her second choice, and so on, and any candidate receiving the most points is a winner.

What is it about scoring-protocol election systems that makes some have the desirable property of being NP-complete to manipulate, while others can be manipulated in polynomial time? We find the complete, dichotomizing answer: Diversity of dislike. Every scoring-protocol election system having two or more point values assigned to candidates other than the favorite—i.e., having $||\{\alpha_i \, | \, 2 \leq i \leq m\}|| \geq 2$—is NP-complete to manipulate. Every other scoring-protocol election system can be manipulated in polynomial time. In effect, we show that—other than trivial systems (where all candidates alway tie), plurality voting, and plurality voting's transparently disguised translations—*every* scoring-protocol election system is NP-complete to manipulate.

## 1 Introduction

Voting provides is a broad framework for preference aggregation (see, e.g., [RO73,NR76, AB00,BF02] for general background). Voting protocols are increasingly becoming natural to consider not just in "human" settings, such as electing senators, but also in computational settings, such as aggregating ranks of web pages and avoiding spam results from web searches [DKNS01,FKS03], collaborative filtering [PHG00], and planning in automated multiagent systems [ER93,ER91].

One central vulnerability of voting systems is that voters may not vote sincerely. The universality of this worry became devastatingly clear in the 1970s through the famous

*Supported in part by grant NSF-CCR-0311021. Work done in part while on sabbatical at the University of Rochester and while visiting Julius-Maximilians-Universität.

†Supported in part by grant NSF-CCF-0426761. Work done in part while visiting Julius-Maximilians-Universität.

Gibbard–Satterthwaite Theorem [Gib73,Sat75]: *Every* nondictatorial (i.e., no one voter controls the outcome regardless of the preferences of the other voters) voting system among three or more candidates is vulnerable to manipulation (i.e., has cases where some voter does better by voting insincerely).

In "The Computational Complexity Difficulty of Manipulating an Election," Bartholdi, Tovey, and Trick ([BTT89a], see also [BO91]) suggested that in light of the impossibility of having an election system remove any temptation to manipulate, one might at least seek election systems that were *computationally resistant* to manipulation, i.e., in which a voter's task in deciding how to vote strategically was complex (e.g., NP-complete). They studied unweighted voters in elections in which the number of candidates is unbounded and there is exactly one strategic voter. For a number of specific election systems, they proved P results and NP-completeness results.

Conitzer and Sandholm, and Conitzer, Lang, and Sandholm ([CS02,CLS03], henceforth referred to collectively as "CS/CLS") proposed and studied a somewhat different model: Voters are weighted, and manipulation is being attempted not by a single voter but by a coalition of strategic voters. Weighted voters are very natural in computational settings, but also often occur in human settings. For example, in the past, all counties in New York State other than the five of New York City used weighted voting systems—with different localities having differing weights typically due to differing sizes—for their governance [Mal04] and, though it is (due to the issue of "unfaithful electors" and the fact that a few states don't allocate their electors as a block) not a perfect example, the Electoral College, which chooses the American president, is akin to a weighted voting system.

Studying weighted voters has a nice side benefit. As CS/CLS note, the NP-hardness results of Bartholdi, Tovey, and Trick [BTT89a] inherently (unless P = NP) depend—due to the unweighted voter model—on the fact that the number of candidates is unbounded. In contrast, in the CS/CLS model, NP-completeness results can meaningfully be obtained even for elections limited to small, fixed numbers of candidates. CS/CLS study some specific voting systems and obtain for them polynomial-time manipulation algorithms and NP-hardness results. For the fixed scoring-protocol systems they studied—plurality, Borda, and veto—they found the number of candidates (if any) where the manipulation complexity changes from P to NP-complete (respectively, never, 3, and 3). Our dichotomy theorem finds the broader pattern of which those are pieces.

That is, the goal of the present paper is to shift the analysis away from specific systems, and instead to ask *What is it about a voting system that makes its manipulation problem easy or hard?* We completely answer this question for the broad range of voting systems, known as *scoring protocols* or *scoring rules* (see the handbook article [BF02]), which have been widely studied in social choice theory.

An ($m$-candidate) scoring-protocol election system is defined by a vector

$(\alpha_1, \alpha_2, \ldots, \alpha_m)$, $\alpha_1 \geq \alpha_2 \geq \cdots \geq \alpha_m$, of integers. Throughout this paper, each voter orders the candidate completely and without ties. Formally, each voter has a preference order that is irreflexive and antisymmetric (that is, each voter has strict preferences over the candidates), complete (that is, each voter weighs in on all the candidates), and transitive. And each voter contributes $\alpha_1$ points to his/her first choice, $\alpha_2$ points to his/her second choice, and so on. Any candidate obtaining, overall, at least as many points as each other candidate is a *winner*. The most important voting system—plurality voting a.k.a. plurality rule a.k.a. first-past-the-post voting—is a scoring protocol with vector $(1, 0, 0, \ldots, 0)$, as is the very common voting system based on ordinal ballots, Borda count a.k.a. preferential-procedure voting, with vector $(m, m-1, \ldots, 1)$, and as are veto voting a.k.a. inverse plurality rule a.k.a. antiplurality rule (with vector $(1, 1, \ldots, 1, 0)$), $k$-approval voting (with vector $(1, \ldots, 1, 0, \ldots, 0)$, starting with $k$ 1's), half-approval voting, in which half the $m$ candidates—the first $\lfloor m/2 \rfloor$ of them—in each voter's ordering get 1 point each from that voter and the rest get zero points each from that voter, and many more.[1] We sometimes take our $m$ candidates to be implicitly named 1, 2, …, $m$.

For a fixed scoring-protocol election system defined by the vector of integers $\alpha = (\alpha_1, \alpha_2, \ldots, \alpha_m)$, $\alpha_1 \geq \alpha_2 \geq \cdots \geq \alpha_m$, following CS/CLS, the (Weighted) Manipulation Problem, which we will denote $\text{MP}_\alpha$, is defined as:

(Weighted) Manipulation Problem, $\text{MP}_\alpha$

**Given:** A set $S$ of weighted voters with preferences over (the same) $m$ candidates, the weights for a set $T$ of voters, and a preferred candidate $p$ from among the $m$ candidates.

**Question:** Is there a way to cast the votes in $T$ such that $p$ wins the election with respect to $\alpha$?

(As is standard, we assume the codings are the natural ones. $S$ is input as a list, each element of which contains a weight written in binary and a preference expressed as a permuted list of the set $\{1, 2, \ldots, m\}$. $T$ is input as a list, each element being a weight written in binary.)

For manipulating to create a *unique* (i.e., sole) winner, the manipulation problem is analogously defined.

Unique (Weighted) Manipulation Problem, $\text{UMP}_\alpha$

**Given:** A set $S$ of weighted voters with preferences over (the same) $m$ candidates, the weights for a set $T$ of voters, and a preferred candidate $p$ from among the $m$ candidates.

**Question:** Is there a way to cast the votes in $T$ such that $p$ is the unique winner of the election with respect to $\alpha$?

---
[1]Some complex-seeming systems also turn out to have very nice properties, e.g., in the case of half-approval, regarding asymptotics of manipulating coalition size [PS].

We prove that for any scoring-protocol election system, the manipulation problem in both the "winner" and the "unique winner" versions is NP-complete exactly if $||\{\alpha_i \mid 2 \leq i \leq m\}|| \geq 2$. And for every scoring-protocol election system violating that condition, the manipulation problem in both the "winner" and the "unique winner" versions is in P. That is, we provide a dichotomy theorem that fully classifies the complexity of manipulation of scoring-protocol election systems. In particular, *every* scoring-protocol election system—other than trivial elections in which all candidates always tie, plurality voting, and an infinite class of systems that are just transparently scaled/transformed clones of plurality voting (see the proof of Theorem 2.1)—is NP-complete to manipulate. (And our NP-completeness results apply to the actual, untainted election systems. We mention in passing that there are papers that fight manipulation in a totally different way: by changing election systems via adding an initial elimination round [CS03, EL].)

The above dichotomy theorem is for elections over fixed numbers of candidates, as that is the standard model for scoring protocols. However, we note that our dichotomy theorem extends naturally to a dichotomy theorem covering unbounded numbers of candidates.

## 2 Main Result: Dichotomy Theorem for Fixed Number of Candidates

**Theorem 2.1** *For each $m$ and each scoring protocol $\alpha = (\alpha_1, \ldots, \alpha_m)$, $\mathrm{MP}_\alpha$ and $\mathrm{UMP}_\alpha$ are in P if $\alpha_2 = \alpha_3 = \cdots = \alpha_m$, and are NP-complete in all other cases.*

In the proof of Theorem 2.1, we will use the following observation, which allows us to put scoring protocols into a normal form.

**Observation 2.2** *Let $\alpha = (\alpha_1, \alpha_2, \alpha_3, \ldots, \alpha_m)$ be a scoring protocol. Then, for all sets of voters $S$ and all candidates $p$, the following hold.*

- *For all integers $k$, $\alpha + k = (\alpha_1 + k, \alpha_2 + k, \alpha_3 + k, \ldots, \alpha_m + k)$ is a scoring protocol and $p$ is a winner (unique winner) with respect to $S$ and $\alpha$ if and only if $p$ is a winner (unique winner) with respect to $S$ and $\alpha + k$.*

- *For all positive integers $k$, $k\alpha = (k\alpha_1, k\alpha_2, k\alpha_3, \ldots, k\alpha_m)$ is a scoring protocol and $p$ is a winner (unique winner) with respect to $S$ and $\alpha$ if and only if $p$ is a winner (unique winner) with respect to $S$ and $k\alpha$.*

**Proof**  Immediate from the observation that for all candidates $c$, the score of $c$ with respect to $S$ and $\alpha + k$ is equal to (the score of $c$ with respect to $S$ and $\alpha$) $+ k||S||$, and the score of $c$ with respect to $S$ and $k\alpha$ is equal to $k$ times (the score of $c$ with respect to $S$ and $\alpha$). ❑

**Proof of Theorem 2.1**  The "in P" part of this theorem is easy to see. If $\alpha_1 = \alpha_2 = \alpha_3 = \cdots = \alpha_m$, then all candidates are always tied, and so are all winners if ties are allowed, and are unique winners if and only if $m = 1$. If $\alpha_1 > \alpha_2 = \alpha_3 = \cdots = \alpha_m$, then, by Observation 2.2, winning in this protocol is equivalent to winning in the protocol $(1, 0, 0, \ldots, 0)$, i.e., plurality, for which the manipulation problem is in P (this is mentioned by CS/CLS and clearly holds, since $p$ is a winner (unique winner) if and only if $p$ is a winner (unique winner) in the election that results when setting every vote in $T$ to $p > \cdots$; we mention in passing that this also works for an unbounded number of candidates).

It is immediate that $MP_\alpha$ and $UMP_\alpha$ are in NP for all scoring protocols (simply guess the votes in $T$ and evaluate the resulting election; it is clear that evaluating such an election can be done in polynomial time). It remains to show that if it is not the case that $\alpha_2 = \alpha_3 = \cdots = \alpha_m$, then $MP_\alpha$ and $UMP_\alpha$ are NP-hard. We will prove this by a reduction from the well-known NP-complete problem PARTITION:

**Given:** $n$ ($n \geq 1$) positive integers $k_1, \ldots, k_n$ that sum to $2K$.

**Question:** Is there a subset of the integers that sums to $K$?

For the remainder of the proof, fix a scoring protocol $\alpha = (\alpha_1, \ldots, \alpha_m)$ such that it is not the case that $\alpha_2 = \alpha_3 = \cdots = \alpha_m$. We will assume without loss of generality (see Observation 2.2) that $\alpha_m = 0$.

Given a $k_1, \ldots, k_n, K$ whose membership in PARTITION we wish to test, we will in polynomial time construct a set of voters $S$, the weights for voters in $T$, and a preferred candidate $p$ such that

- If there exists a subset of $k_1, \ldots, k_n$ that sums to $K$, then the votes in $T$ can be cast such that $p$ becomes a unique winner (with respect to $S \cup T$).

- If the votes in $T$ can be cast in such a way that $p$ becomes a winner, then there exists a subset of $k_1, \ldots, k_n$ that sums to $K$.

Note that this shows NP-hardness for $MP_\alpha$ and $UMP_\alpha$.

Let $\ell = ||\{i \mid 1 \leq i \leq m \text{ and } \alpha_i = 0\}|| - 1$. Since $\alpha_m = 0$, $\ell \geq 0$. Let $r = m - \ell - 3$. Since $\alpha_2 \neq 0$, $r \geq 0$. Our election consists of the following $m$ candidates: $p$, $a$, $b$, $c_1, \ldots, c_\ell$ and $d_1, \ldots, d_r$.

The set of voters $S$ will consist of two parts, $S = S_1 \cup S_2$. $S_1$ will ensure that the reduction works if we look only at candidates $a$, $b$, $p$, and $c_1, \ldots, c_\ell$. $S_2$ will ensure that there is no interference from the padding candidates $d_1, \ldots, d_r$.

First suppose that $\ell \neq 1$ (we will need a slightly different construction for $\ell = 1$, which we will describe after the proof of the $\ell \neq 1$ case).

**The $\ell \neq 1$ case**

$S_1$ consists of the following voters:

- One voter of weight $2K(2\alpha_1 - \alpha_{r+2}) - 1$ of the form
$$a > d_1 > \cdots > d_r > b > p > c_1 > \cdots > c_\ell.^2$$

- One voter of weight $2K(2\alpha_1 - \alpha_{r+2}) - 1$ of the form
$$b > d_1 > \cdots > d_r > a > p > c_1 > \cdots > c_\ell.$$

- For each $i$, $1 \leq i \leq \ell$, one voter of weight $4\alpha_1 K - 1$ of the form
$$c_i > d_1 > \cdots > d_r > c_{1+(i \bmod \ell)} > a > b > p > \cdots,$$

  where the final "$\cdots$" means that the remaining candidates follow in some arbitrary order.

For $V$ a set of voters and $c$ a candidate, define $score_V(c)$ as the score of candidate $c$ with respect to $V$ (and $\alpha$).

It is easy to verify that the scores of the candidates with respect to $S_1$ are as follows:

1. $score_{S_1}(p) = 0$.
2. $score_{S_1}(a) = score_{S_1}(b) = (2K(2\alpha_1 - \alpha_{r+2}) - 1)(\alpha_1 + \alpha_{r+2})$.
3. $score_{S_1}(c_i) = (4\alpha_1 K - 1)(\alpha_1 + \alpha_{r+2})$, for each $1 \leq i \leq \ell$.
4. $score_{S_1}(d_i) \leq score_{S_1}(d_1) = (2(2K(2\alpha_1 - \alpha_{r+2}) - 1) + \ell(4\alpha_1 K - 1))\alpha_2$, for each $1 \leq i \leq r$.

Note that the candidates $d_i$ have (potentially) very high scores. We will define a set of voters $S_2$ that will ensure that the $d_i$'s will not interfere with $p$ winning the election. In order to do so, we will use the following claim. (Note that our scoring protocol is fixed, so $\ell$ and $r$ are fixed. So the "polynomial time" below is in the context of these fixed $\ell$ and $r$, with $s$ our variable.)

**Claim 2.3** *For every positive integer $s$, there exists a set of voters (each having a weight and a preference order) $S_2$ over candidates $\{p, a, b, c_1, \ldots, c_\ell, d_1, \ldots, d_r\}$ such that $score_{S_2}(p) = score_{S_2}(x)$ for all $x \in \{a, b, c_1, \ldots, c_\ell\}$, and $score_{S_2}(d_i) + s \leq score_{S_2}(p)$ for all $1 \leq i \leq r$. From $s$, in polynomial time (in $|s|$) we may compute such an $S_2$.*

---
[2]It would be tempting to instead say "$2K(2\alpha_1 - \alpha_{r+2}) - 1$ voters of weight 1 of the form $a > d_1 > \cdots > d_r > b > p > c_1 > \cdots > c_\ell$." But note that that is invalid as such a set of voters takes exponential space (in the length of the input), since the integers in PARTITION are in binary. (The unary version of PARTITION is in P.) For this reason, the proofs of [CS02, Theorems 1–4] and [CLS03, Theorems 2 and 7] are technically incorrect, though these proofs can easily be fixed by handling things with high-weight voters as above.

We will prove this claim at the end of the $\ell \neq 1$ case of Theorem 2.1.

Let $S_2$ be a set of voters such that $score_{S_2}(p) = score_{S_2}(x)$ for all $x \in \{a, b, c_1, \ldots, c_\ell\}$, and $score_{S_2}(d_i) + score_{S_1}(d_1) < score_{S_2}(p)$ for all $1 \leq i \leq r$. The existence and polynomial-time computability of such an $S_2$ follow from Claim 2.3, applied with $s = score_{S_1}(d_1) + 1$. Let $S = S_1 \cup S_2$, and set the weights of $T$ to

$$2k_1(\alpha_1 + \alpha_{r+2}), 2k_2(\alpha_1 + \alpha_{r+2}), \ldots, 2k_n(\alpha_1 + \alpha_{r+2}).$$

We will show that we have obtained the desired reduction.

First suppose that there exists a subset of $k_1, \ldots, k_n$ that sums to $K$. We need to show that we can cast the votes in $T$ in such a way that $p$ becomes the unique winner (with respect to $S \cup T$).

Let $I \subseteq \{1, \ldots, n\}$ be such that $\sum_{i \in I} k_i = K$. There are $n$ voters in $T$, and we will view these voters as numbered from 1 to $n$ in such a way that the $i$th voter in $T$ has weight $2k_i(\alpha_1 + \alpha_{r+2})$. For every $i$, $1 \leq i \leq n$, set the vote of the $i$th voter in $T$ to

$$p > d_1 > \cdots > d_r > a > b > c_1 > \cdots > c_\ell \text{ if } i \in I$$

and to

$$p > d_1 > \cdots > d_r > b > a > c_1 > \cdots > c_\ell \text{ if } i \notin I.$$

This is in the current setting equivalent to having one voter of weight $2K(\alpha_1 + \alpha_{r+2})$ of the form

$$p > d_1 > \cdots > d_r > a > b > c_1 > \cdots > c_\ell$$

and one voter of weight $2K(\alpha_1 + \alpha_{r+2})$ of the form

$$p > d_1 > \cdots > d_r > b > a > c_1 > \cdots > c_\ell.$$

Let $s_2 = score_{S_2}(p)$. Recall that $score_{S_2}(a) = s_2$, $score_{S_2}(b) = s_2$, and $score_{S_2}(c_i) = s_2$ for all $1 \leq i \leq \ell$. We obtain the following scores (with respect to $S \cup T$).

1. $score(p) = s_2 + 4\alpha_1 K(\alpha_1 + \alpha_{r+2})$.

2. $score(a) = score(b) = (2K(2\alpha_1 - \alpha_{r+2}) - 1)(\alpha_1 + \alpha_{r+2}) + s_2 + 2K(\alpha_1 + \alpha_{r+2})\alpha_{r+2} = s_2 + (2K(2\alpha_1 - \alpha_{r+2}) - 1 + 2K\alpha_{r+2})(\alpha_1 + \alpha_{r+2}) = s_2 + (4K\alpha_1 - 1)(\alpha_1 + \alpha_{r+2})$, for each $1 \leq i \leq \ell$. This is less than $score(p)$.

3. $score(c_i) = (4\alpha_1 K - 1)(\alpha_1 + \alpha_{r+2}) + s_2$, for each $1 \leq i \leq \ell$. This is less than $score(p)$.

4. For each $1 \leq i \leq r$, $score(d_i) \leq score(d_1) = score_S(d_1) + score_T(d_1) < score_S(p) + score_T(p) = score(p)$ (the "$<$" is due to our choice of $S_2$ and the preferences of the voters in $T$).

It follows that $p$ is the unique winner of this election, as required.

For the converse, suppose the votes in $T$ are cast in such a way that $p$ is a winner of the election. Without loss of generality, assume that $p$ is the most preferred candidate in the preference order of each voter in $T$. Then $score(p) = s_2 + 4\alpha_1 K(\alpha_1 + \alpha_{r+2})$.

We will now show that, for all $i$, $1 \leq i \leq \ell$, $score_T(c_i) = 0$. Suppose that for some $i$ it holds that $score_T(c_i) > 0$. Since all weights of $T$ are multiples of $2(\alpha_1+\alpha_{r+2})$, it follows that $score_T(c_i) \geq 2(\alpha_1+\alpha_{r+2})$. But then $score(c_i) \geq (4\alpha_1 K-1)(\alpha_1+\alpha_{r+2})+s_2+2(\alpha_1+\alpha_{r+2}) = s_2+(4\alpha_1 K+1)(\alpha_1+\alpha_{r+2}) > score(p)$. This contradicts the assumption that $p$ is a winner.

Recall that there are only $\ell + 1$ 0's in $\alpha$. Since $score_T(c_i) = 0$ for $1 \leq i \leq \ell$, each voter in $T$ must have at least one of $a$ or $b$ somewhere between (inclusively) 1st and $(r + 2)$nd in his/her preference order. So it follows that

$$score_T(a) + score_T(b) \geq 4K(\alpha_1 + \alpha_{r+2})\alpha_{r+2}.$$

Suppose that $score_T(a) > 2K(\alpha_1 + \alpha_{r+2})\alpha_{r+2}$. Since all weights of $T$ are multiples of $2(\alpha_1 + \alpha_{r+2})$, it follows that $score_T(a) \geq 2K(\alpha_1+\alpha_{r+2})\alpha_{r+2} + 2(\alpha_1 + \alpha_{r+2}) = (2K\alpha_{r+2}+2)(\alpha_1+\alpha_{r+2})$. Then (keeping in mind our choice of $S_2$) $score(a) \geq (2K(2\alpha_1-\alpha_{r+2})-1)(\alpha_1+\alpha_{r+2})+s_2+(2K\alpha_{r+2}+2)(\alpha_1+\alpha_{r+2}) = s_2+(2K(2\alpha_1-\alpha_{r+2})+1+2K\alpha_{r+2})(\alpha_1+\alpha_{r+2}) = s_2+(4K\alpha_1+1)(\alpha_1+\alpha_{r+2})$. This is greater than $score(p)$. This contradicts the assumption that $p$ is a winner.

Since $a$ and $b$ are completely symmetric, it follows that $score_T(a) = score_T(b) = 2K(\alpha_1+\alpha_{r+2})\alpha_{r+2}$. But then the weights of those voters in $T$ who prefer $a$ to $b$ sum to $2K(\alpha_1+\alpha_{r+2})$. This implies that there is a subset $I \subseteq \{1,\ldots,n\}$ such that $\sum_{i\in I} 2k_i(\alpha_1+\alpha_{r+2}) = 2K(\alpha_1+\alpha_{r+2})$. It follows that $\sum_{i\in I} k_i = K$.

To finish the proof of the $\ell \neq 1$ case, it remains to prove Claim 2.3, i.e., to construct in time polynomial in $|s|$ a set of voters $S_2$ over candidates $\{p, a, b, c_1, \ldots, c_\ell, d_1, \ldots, d_r\}$ such that $score_{S_2}(p) = score_{S_2}(x)$ for all $x \in \{a, b, c_1, \ldots, c_\ell\}$, and $score_{S_2}(d_i) + s \leq score_{S_2}(p)$ for all $1 \leq i \leq r$.

**Proof of Claim 2.3** If $r = 0$, we simply take $S_2 = \emptyset$. So, assume that $r > 0$. In the construction of $S_2$, we will use cyclic shifts of preference orders in order to make sure that certain candidates tie. We introduce the following notation.

For a preference order $a_0 > a_1 > \cdots > a_{k-1}$, and integer $t$, $[a_0 > a_1 > \cdots > a_{k-1}]_{t\rightarrow}$ denotes the preference order that results after a cyclic shift of $t$ positions to the right, i.e., the preference order $a_{t \bmod k} > a_{(t+1) \bmod k} > \cdots > a_{(t+k-1) \bmod k}$.

$S_2$ is defined as follows.

- For every $i$, $0 \leq i \leq \ell + 2$, and every $j$, $0 \leq j \leq r - 1$, a voter of weight $s$ of the form:

$$[a > b > p > c_1 > \cdots > c_\ell]_{i\rightarrow} > [d_1 > \cdots > d_r]_{j\rightarrow}.$$

So, $S_2$ consists of $(\ell+3)r$ voters, each of weight $s$. Clearly, for each $1 \le i \le \ell$, $score_{S_2}(p) = score_{S_2}(a) = score_{S_2}(b) = score_{S_2}(c_i)$, as required.

It remains to show that, for each $1 \le i \le r$, $score_{S_2}(d_i) + s \le score_{S_2}(p)$. Note that $score_{S_2}(p) = rs(\alpha_1 + \cdots + \alpha_{\ell+3})$ and $score_{S_2}(d_i) = (\ell+3)s(\alpha_{\ell+3+1} + \cdots + \alpha_m)$. In light of this, $score_{S_2}(d_i) + s \le score_{S_2}(p)$ will clearly follow if we can establish that $r(\alpha_1 + \cdots + \alpha_{\ell+3}) > (\ell+3)(\alpha_{\ell+3+1} + \cdots + \alpha_m)$. But that indeed holds, since $r(\alpha_1 + \cdots + \alpha_{\ell+3}) \ge r(\ell+3)\alpha_{\ell+3} = (m - \ell - 3)\alpha_{\ell+3}(\ell+3) \ge (\alpha_{\ell+3+1} + \cdots + \alpha_m)(\ell+3)$, and the leftmost "$\ge$" is strict whenever $\alpha_{\ell+3} = 0$ and the rightmost "$\ge$" is strict whenever $\alpha_{\ell+3} \ne 0$.

So, we have $score_{S_2}(p) - score_{S_2}(d_i) \ge s$. ❑ Claim 2.3

This completes the proof of Theorem 2.1 for the case that $\ell \ne 1$. Note that the construction given above does not work for $\ell = 1$. We will now modify the previous construction to get a construction for the $\ell = 1$ case.

**The $\ell = 1$ case**

$S_1$ consists of the following voters:

- One voter of weight $3K(2\alpha_1 - \alpha_{r+2}) - 1$ of the form

$$a > d_1 > \cdots > d_r > b > p > c_1.$$

- One voter of weight $3K(2\alpha_1 - \alpha_{r+2}) - 1$ of the form

$$b > d_1 > \cdots > d_r > a > p > c_1.$$

- The following six voters, each of weight $3\alpha_1 K - 1$:

$$\begin{aligned}
c_1 > d_1 > \cdots > d_r > a > b > p, \\
a > d_1 > \cdots > d_r > c_1 > b > p, \\
c_1 > d_1 > \cdots > d_r > b > a > p, \\
b > d_1 > \cdots > d_r > c_1 > a > p, \\
c_1 > d_1 > \cdots > d_r > p > a > b, \text{ and} \\
p > d_1 > \cdots > d_r > c_1 > a > b.
\end{aligned}$$

It is easy to verify that the scores of the candidates with respect to $S_1$ are as follows:

1. $score_{S_1}(p) = (3\alpha_1 K - 1)(\alpha_1 + \alpha_{r+2})$.

2. $score_{S_1}(a) = score_{S_1}(b) = (3K(2\alpha_1 - \alpha_{r+2}) - 1)(\alpha_1 + \alpha_{r+2}) + score_{S_1}(p)$.

3. $score_{S_1}(c_1) = 2(3\alpha_1 K - 1)(\alpha_1 + \alpha_{r+2}) + score_{S_1}(p)$.

4. $score_{S_1}(d_i) \leq score_{S_1}(d_1) = (2(3K(2\alpha_1 - \alpha_{r+2}) - 1) + 6(3\alpha_1 K - 1))\alpha_2$, for each $1 \leq i \leq r$.

Let $S_2$ be a set of voters such that $score_{S_2}(p) = score_{S_2}(x)$ for all $x \in \{a, b, c_1\}$, and $score_{S_2}(d_i) + score_{S_1}(d_1) \leq score_{S_2}(p)$ for all $1 \leq i \leq r$. The existence and polynomial-time computability of such an $S_2$ follow from Claim 2.3, applied with $s = score_{S_1}(d_1)$.

Let $S = S_1 \cup S_2$, and set the weights of $T$ to

$$3k_1(\alpha_1 + \alpha_{r+2}), 3k_2(\alpha_1 + \alpha_{r+2}), \ldots, 3k_n(\alpha_1 + \alpha_{r+2}).$$

We will show that that we have obtained the desired reduction.

First suppose that there exists a subset of $k_1, \ldots, k_n$ that sums to $K$. We need to show that we can cast the votes in $T$ in such a way that $p$ becomes the unique winner (with respect to $S \cup T$).

Let $I \subseteq \{1, \ldots, n\}$ be such that $\sum_{i \in I} k_i = K$. There are $n$ voters in $T$, and we will view these voters as numbered from 1 to $n$ in such a way that the $i$th voter in $T$ has weight $3k_i(\alpha_1 + \alpha_{r+2})$. For every $i$, $1 \leq i \leq n$, set the vote of the $i$th voter in $T$ to

$$p > d_1 > \cdots > d_r > a > b > c_1 \text{ if } i \in I$$

and to

$$p > d_1 > \cdots > d_r > b > a > c_1 \text{ if } i \notin I.$$

This is in the current setting equivalent to having one voter of weight $3K(\alpha_1 + \alpha_{r+2})$ of the form

$$p > d_1 > \cdots > d_r > a > b > c_1$$

and one voter of weight $3K(\alpha_1 + \alpha_{r+2})$ of the form

$$p > d_1 > \cdots > d_r > b > a > c_1.$$

Let $s_1 = score_{S_1}(p)$ and let $s_2 = score_{S_2}(p)$. Recall that, by our choice of $S_2$, $score_{S_2}(a) = s_2$, $score_{S_2}(b) = s_2$, and $score_{S_2}(c_1) = s_2$. We obtain the following scores (with respect to $S \cup T$).

1. $score(p) = s_1 + s_2 + 6\alpha_1 K(\alpha_1 + \alpha_{r+2})$.

2. $score(a) = score(b) = s_1 + (3K(2\alpha_1 - \alpha_{r+2}) - 1)(\alpha_1 + \alpha_{r+2}) + s_2 + 3K(\alpha_1 + \alpha_{r+2})\alpha_{r+2} = s_1 + s_2 + (3K(2\alpha_1 - \alpha_{r+2}) - 1 + 3K\alpha_{r+2})(\alpha_1 + \alpha_{r+2}) = s_1 + s_2 + (6K\alpha_1 - 1)(\alpha_1 + \alpha_{r+2})$. This is less than $score(p)$.

3. $score(c_1) = (6\alpha_1 K - 2)(\alpha_1 + \alpha_{r+2}) + s_1 + s_2$. This is less than $score(p)$.

4. For each $1 \leq i \leq r$, $score(d_i) = score_{S_1}(d_i) + score_{S_2}(d_i) + score_T(d_i) \leq score_{S_1}(d_1) + score_{S_2}(d_i) + score_T(d_1) \leq score_{S_2}(p) + score_T(p) < score(p)$ (the rightmost "$\leq$" is due to our choice of $S_2$ and the preferences of the voters in $T$).

It follows that $p$ is the unique winner of this election, as required.

For the converse, suppose the votes in $T$ are cast in such a way that $p$ is a winner of the election. Without loss of generality, assume that $p$ is the most preferred candidate in the preference order of each voter in $T$. Then $score(p) = s_1 + s_2 + 6\alpha_1 K(\alpha_1 + \alpha_{r+2})$.

We will now show that $score_T(c_1) = 0$. Suppose that $score_T(c_1) > 0$. Since all weights of $T$ are multiples of $3(\alpha_1 + \alpha_{r+2})$, it follows that $score_T(c_1) \geq 3(\alpha_1 + \alpha_{r+2})$. But then $score(c_1) \geq (6\alpha_1 K - 2)(\alpha_1 + \alpha_{r+2}) + s_1 + s_2 + 3(\alpha_1 + \alpha_{r+2}) = s_1 + s_2 + (6\alpha_1 K + 1)(\alpha_1 + \alpha_{r+2}) > score(p)$. This contradicts the assumption that $p$ is a winner.

Recall that there are only two 0's in $\alpha$. Since $score_T(c_1) = 0$, each voter in $T$ must have at least one of $a$ or $b$ somewhere between (inclusively) 1st and $(r+2)$nd in his/her preference order, so it follows that

$$score_T(a) + score_T(b) \geq 6K(\alpha_1 + \alpha_{r+2})\alpha_{r+2}.$$

Suppose that $score_T(a) > 3K(\alpha_1 + \alpha_{r+2})\alpha_{r+2}$. Since all weights of $T$ are multiples of $3(\alpha_1 + \alpha_{r+2})$, it follows that $score_T(a) \geq 3K(\alpha_1 + \alpha_{r+2})\alpha_{r+2} + 3(\alpha_1 + \alpha_{r+2}) = (3K\alpha_{r+2} + 3)(\alpha_1 + \alpha_{r+2})$. Then (keeping in mind our choice of $S_2$) $score(a) \geq (3K(2\alpha_1 - \alpha_{r+2}) - 1)(\alpha_1 + \alpha_{r+2}) + s_1 + s_2 + (3K\alpha_{r+2} + 3)(\alpha_1 + \alpha_{r+2}) = s_1 + s_2 + (3K(2\alpha_1 - \alpha_{r+2}) + 2 + 3K\alpha_{r+2})(\alpha_1 + \alpha_{r+2}) = s_1 + s_2 + (6K\alpha_1 + 2)(\alpha_1 + \alpha_{r+2})$. This is greater than $score(p)$. This contradicts the assumption that $p$ is a winner.

Since $a$ and $b$ are completely symmetric, it follows that $score_T(a) = score_T(b) = 3K(\alpha_1 + \alpha_{r+2})\alpha_{r+2}$. But then the weights of those voters in $T$ who prefer $a$ to $b$ sum to $3K(\alpha_1 + \alpha_{r+2})$. This implies that there is a subset $I \subseteq \{1, \ldots, n\}$ such that $\sum_{i \in I} 3k_i(\alpha_1 + \alpha_{r+2}) = 3K(\alpha_1 + \alpha_{r+2})$. It follows that $\sum_{i \in I} k_i = K$. ❑

## 3  Unbounded Numbers of Candidates

Our main result, Theorem 2.1, is for the case of fixed numbers of candidates. However, intuitively, some voting systems "feel" as if they should handle arbitrary number of candidates. For example, the family of score vectors $(1)$, $(1, 0)$, $(1, 0, 0)$, ... taken together correctly capture a notion of plurality voting that is not bound to a particular number of candidates. To generalize our main theorem to unbounded numbers of voters, we need a way to specify such families of scoring vectors. We do so in what seems the most natural way: We require there to be an efficient function that generates the score vectors. Let $\mathbb{Z}$ denote the integers. Let us say that a function $f$ is a *general scoring function* if $f$ is polynomial-time computable and, for each $m \geq 1$,

$$f(0^m) = \{(\alpha_1, \alpha_2, \ldots, \alpha_m) \in \mathbb{Z}^m \mid \alpha_1 \geq \alpha_2 \geq \cdots \geq \alpha_m\}.$$

By an election with respect to $f$, we mean exactly the natural notion: If a given input has $m$ candidates, we conduct the election with respect to the scoring rule $f(0^m)$.

So, we may now state our regular and unique problems for the case of scoring protocols that apply to unbounded numbers of candidates. For each general scoring function $f$, we define the following two problems (unlike $\mathrm{MP}_\alpha$ and $\mathrm{UMP}_\alpha$ where $m$ is fixed by $\alpha$, here $m$ is a variable determined by the preferences that are part of the input).

General (Weighted) Manipulation Problem, $\mathrm{GMP}_f$

**Given:** A set $S$ of weighted voters with preferences over (the same) $m$ candidates, the weights for a set $T$ of voters, and a preferred candidate $p$ from among the $m$ candidates.

**Question:** Is there a way to cast the votes in $T$ such that $p$ wins the election with respect to $f(0^m)$?

General Unique (Weighted) Manipulation Problem, $\mathrm{GUMP}_f$

**Given:** A set $S$ of weighted voters with preferences over (the same) $m$ candidates, the weights for a set $T$ of voters, and a preferred candidate $p$ from among the $m$ candidates.

**Question:** Is there a way to cast the votes in $T$ such that $p$ is the unique winner of the election with respect to $f(0^m)$?

It is not hard to see that our main theorem's result extends to these cases as follows. Here, $f(0^m)_i$, $1 \leq i \leq m$, denotes the $i$th component of the arity-$m$ vector $f(0^m)$.

**Theorem 3.1** *For each general scoring function $f$, $\mathrm{GMP}_f$ and $\mathrm{GUMP}_f$ are in P if*

$$(\forall m \geq 1)(\forall k, \ell : 2 \leq k, \ell \leq m)[f(0^m)_k = f(0^m)_\ell],$$

*and are* NP-*complete in all other cases.*

(Just to be clear, what this says regarding the P case is that for all $m$, if one takes the vector output by $f(0^m)$ and chops off its first component, then all the remaining components, if any, are equal to each other.)

Let us explain why this theorem holds for the case of $\mathrm{GMP}_f$. (The $\mathrm{GUMP}_f$ case is analogous.) Fix a general scoring function $f$. Recall that $f$ is polynomial-time computable. If $(\forall m \geq 1)(\forall k, \ell : 2 \leq k, \ell \leq m)[f(0^m)_k = f(0^m)_\ell]$, the same approach that gave P algorithms in the bounded candidate case works here. (Of course, the algorithm will dynamically generate the scoring vector, and then will use the appropriate strategy. It is easy to see that overall this is a polynomial-time algorithm here. This is basically because there were only a finite number of cases underlying the earlier polynomial-time result, so our composite, dynamic algorithm is easily seen to run in polynomial time.) On the other hand, suppose it is not the case that $(\forall m \geq 1)(\forall k, \ell : 2 \leq k, \ell \leq m)[f(0^m)_k = f(0^m)_\ell]$. Then note that there must exist some $m'$ such that the fixed-number-of-candidates problem $\text{MP}_{f(0^{m'})}$ is—via Theorem 2.1—NP-complete. Then clearly our problem $\text{GMP}_f$ is NP-hard, since we have a many-one polynomial-time reduction from the NP-complete problem $\text{MP}_{f(0^{m'})}$ to $\text{GMP}_f$, namely, simply pass forward all the same voters, preferences, and weights. Also, it is directly clear that $\text{GMP}_f$ belongs to NP. Thus, $\text{GMP}_f$ is indeed NP-complete.

## 4  Conclusions

There are some election systems for which even determining the winner is hard (see, e.g., [BTT89b,HHR97,HSV03,RSV03]). Nonetheless, one central goal in the complexity-theoretic study of elections is to find natural, attractive election systems whose winner-evaluation problems are computationally simple but that nonetheless are hard to manipulate.

Scoring protocols all clearly have winner-evaluation problems that are in P. And the dichotomy result of this paper shows that—except for some trivial systems and plurality rule in various guises—*all* scoring-protocol election systems are NP-complete to manipulate.

**Acknowledgments** We thank Klaus Wagner for hosting the visit to Würzburg during which this work was done in part.